\documentclass[twocolumn,pra,superscriptaddress]{revtex4-1}

\usepackage{bm}%
\usepackage{float}
\expandafter\ifx\csname package@font\endcsname\relax\else
 \expandafter\expandafter
 \expandafter\usepackage
 \expandafter\expandafter
 \expandafter{\csname package@font\endcsname}%
\fi

\usepackage{array}
\usepackage{amsmath,amssymb}
\usepackage{MnSymbol}
\usepackage{tikz-feynman,contour} 
\usetikzlibrary{shapes,arrows,positioning,automata,backgrounds,calc,er,patterns}
\usepackage{feynmf}
\tikzfeynmanset{compat=1.0.0} 
\usepackage{graphicx}
\usepackage{mathrsfs}
\usepackage{bm}
\usepackage{mathtools}
\usepackage{epstopdf}
\usepackage{hyperref}

\usepackage{amsmath}
\usepackage{amsfonts}
\usepackage{mathtools}
\usepackage{graphicx}
\usepackage{fixme}
\usepackage{color}

\usepackage{wasysym}

\begin{document}

\title{The Fate of the Upper Polariton: Breakdown of the Quasiparticle Picture in the Continuum }
\author{Yesenia A. Garc\'ia Jomaso}

\author{Brenda Vargas}

\author{David Ley Dominguez}

\author{César L. Ordoñez-Romero}

\author{Hugo A. Lara-García}

\author{Arturo Camacho-Guardian}
\email{acamacho@fisica.unam.mx}

\author{Giuseppe Pirruccio}
\email{pirruccio@fisica.unam.mx}


\affiliation{Instituto de F\'isica, Universidad Nacional Aut\'onoma de M\'exico, Ciudad de M\'exico C.P. 04510, Mexico\looseness=-1}

\date{\today}
\begin{abstract} 

Organic polaritons, strongly hybridized light-matter excitations, have arisen as a platform to device optical interfaces at room temperature. Despite their inherent complexity,  polaritons are commonly regarded as coherent excitations described by Landau's quasiparticle theory. Here, we experimentally and theoretically unravel the role of incoherent matter states by exploring the relevant polariton parameter space. We unveil the fading of the upper polariton at its entrance to a continuum of molecular excitations. This marks the breakdown of the simplistic quasiparticle picture for this branch and the formation of a more complex quantum state beyond Landau’s formalism.

\end{abstract}

\maketitle

Coherent phenomena benefit from the lack of noise~\cite{Zurek2003,Rabitz2000,Viola1999}. Isolation from the environment helps in removing external sources of decoherence and it is routinely applied in quantum technologies: quantum optics~\cite{Orszag2016}, quantum computing~\cite{Alicki2002,Horodecki2009}, and cryptography~\cite{Gisin2002,Ralph1999}. However, this strategy is ineffective against noise arising from internal processes that cannot be disentangled from the system under research.  
Conversely to what it is typically thought, noise is not always detrimental and can drive unexpected phenomena at all scales~\cite{noise3,noise1,noise4,noise2}. 
At the nanoscale, the presence of noise or disorder in complex systems gives rise to random lasers~\cite{Albada1985,John1987,Lawandy1994}, Anderson localization~\cite{Anderson1958}, and enhanced energy transport~\cite{Long2008}. Stochastic resonances~\cite{Gammaitoni1998}, Purcell effect~\cite{Purcell1946}, Casimir forces~\cite{Casimir1948}, and radiative heat transfer~\cite{polder,biehs} are other important manifestations of noise-driven physics. 
Frequency noise is epitomized by the broadening of a resonance linewidth. 
Even though only white noise truly embodies a continuum, a sufficiently wide frequency window of available competing states effectively plays the role of a continuum. 
\begin{figure}[h!]
\centering
    \includegraphics[width=0.9\columnwidth]{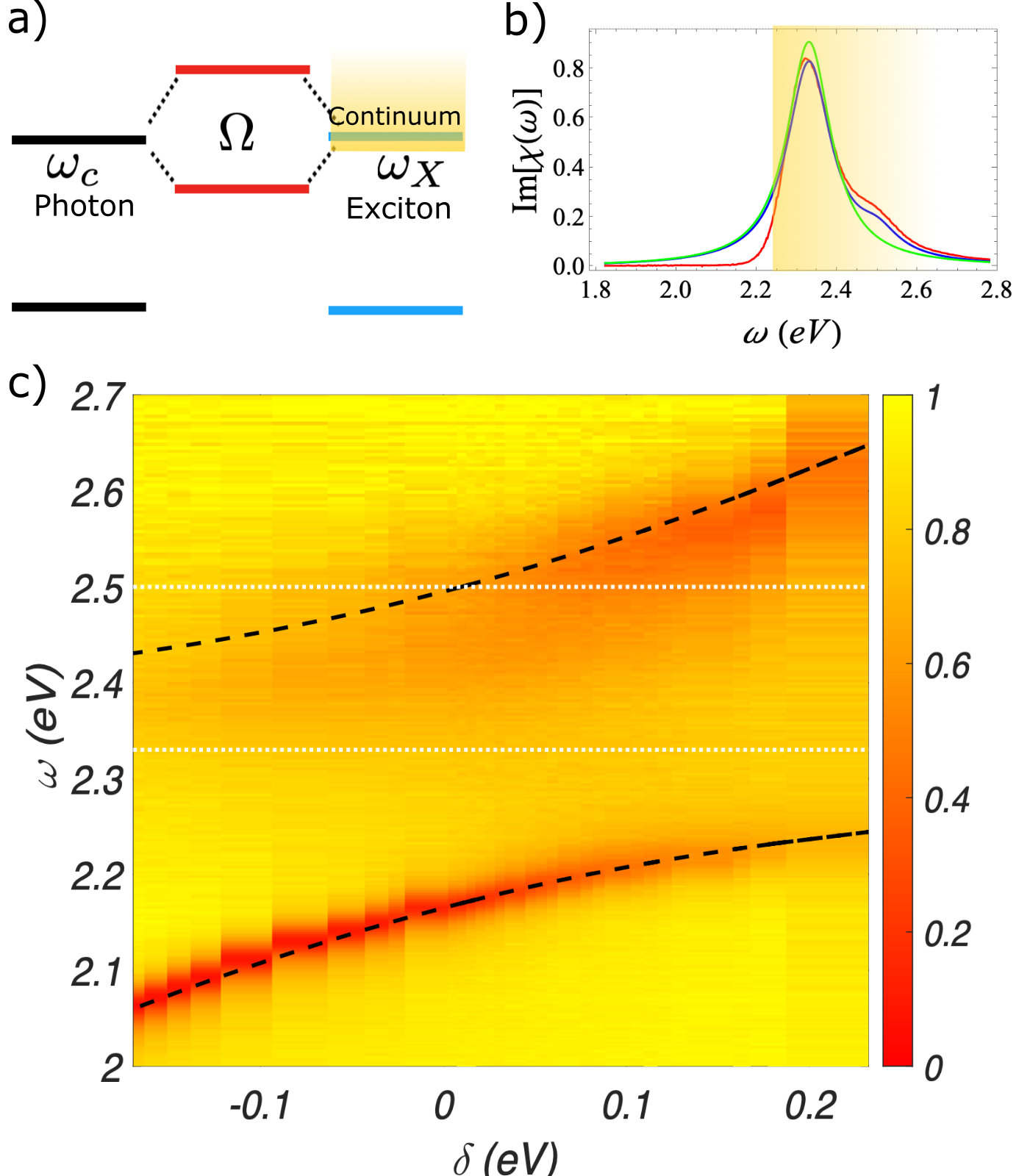}
    \caption{ (a) Relevant energy diagrams for the photon and exciton fields, the latter being embedded in a continuum of incoherent excitations. (b) Measured absorption spectrum of the active layer expressed as $\text{Im}(\chi(\omega))$ (red curve). Blue curve corresponds to  the theoretical modelling of the absorption, while the green one to that of a single exciton peak with a standard broadening $\gamma_X.$ c) Experimental reflectance spectra at normal incidence as a function of the cavity detuning. Dashed black lines correspond to the LP and UP energies while the white lines are guidelines for the continuum broadening of the Frenkel exciton.} 
    \label{Fig1}
\end{figure}

A novel and intriguing class of coherent excitations are cavity polaritons, which result from the strong coupling between confined photons and matter excitations~\cite{Hopfield1958,PEKAR1958}. Polaritons are usually described within Landau's quasiparticle theory, which stands  nowadays as one of the most powerful concepts in physics~\cite{Landau1950a,Landau1950b,Baym1991}. Describing the linear hybridisation of light and matter in terms of polaritons reduces dramatically the complexity of the system. The realm of Landau's theory extends to interactions and allows for the study of effective photon-photon interactions~\cite{Vladimirova2010,Takemura2014,Tan2020,Schwartz2021}. Dressed by their environments, a new quasiparticle coined polaron-polariton can also summarise the many-body complexity of the system~\cite{Sidler2017,Levinsen2019,Navadeh2019,Bastarrachea2019,Emmanuele2020}. However, intrinsic or extrinsic processes can lead to the breakdown of the polaritons picture where their description cannot longer be captured by the quasiparticle properties. Organic polaritons feature complicated inherent processes result of the large number of internal degrees of freedom~\cite{Kena2008,Herrera2017,Keeling2020,Liu2020,Sanchex2022}. Understanding the extent of the quasiparticle theory remains as an experimental and theoretical challenge in these systems.

Motivated by the rapid progress and the necessity to deepen the understanding of organic polaritons, in this Letter we propose a strongly coupled optical cavity embedding a dye-doped polymer to experimentally unravel the role of a continuum of matter excitations on the quasiparticle character of exciton-polaritons.  We demonstrate that as a consequence of the underlying intricate energy level structure that extends beyond the simplified two-level model for the Frenkel exciton, a continuum of matter excitations dramatically affects the coherence of the polaritons. The experimental control of the energy and momentum of the polaritons allows us to gradually tune one of the polariton branches inside the matter continuum and reveal the fading of the quasiparticle picture. We sustained our experimental observations by a general field theory which, in very good accordance with the experiment, confirms the validity and failure of the quasiparticle picture for the polariton branches across the relevant energy-momentum parameter space. We anticipate that the breakdown of the quasiparticle picture may have important consequences on the polariton ultrafast dynamics and may underpin important cavity-mediated phenomena such as condensation~\cite{condensate1,condensate2}, polariton chemistry~\cite{joel}, singlet fission~\cite{singlet}, vibrational strong-coupling~\cite{ebbesen}, ultra-strong coupling~\cite{usc1,usc2}, temperature-activated delayed fluorescence and phosphorescence~\cite{erb1}.

{\it System.-} Our nanocavity consists of two Ag mirrors sandwiching a thin active layer of ErB molecules embedded in a polyvinyl alcohol (PVA) matrix. A 0.5 M concentration of ErB ensures the system is in the strong light-matter coupling regime. The cavity is fabricated by first sputtering a Ag mirror of thickness $300~\text{nm}$ on top of a glass substrate. The active layer of nominal thickness $L_c$ is deposited by spin-coating and, finally, a second Ag mirror of thickness $25~\text{nm}$ is sputtered on top of it. For all practical means, the first mirror can be treated as semi-infinite. $L_c$ is chosen such that the energy of the fundamental cavity mode lies close to the exciton one, i.e., $L_c$ is around 160 nm depending on the cavity detuning to be studied. 

The optical response of the active medium can be described by its optical susceptibility $\chi(\omega).$ The absorption of the layer, captured by $\text{Im}\chi(\omega)$, is illustrated with a red curve in Fig.~\ref{Fig1}(b) (see \cite{SM}). The absorption spectrum exhibits non-trivial features associated with vibrational bands, exciton interactions, anisotropies, among others~\cite{Stomphorst2001}. It consists of the main peak centered at $\omega_X\approx 2.33~\text{eV}$ associated with the principal $S_0\rightarrow S_1$ transition and an overlapping second mode that gives rise to a bump close to $\omega\approx 2.5~\text{eV}.$ The non-trivial broadening of both results in a continuous range of excitations, which we simply refer to as {\it matter continuum} (see Fig.~\ref{Fig1}(b)). Rather than focusing on its microscopic origin, which lies beyond the scope of this Letter, here we unveil the effects of this continuum on the quasiparticle character of the exciton-polaritons.

{\it Cavity polaritons.-} The collective coupling of the organic molecules to the fundamental cavity mode give rises to two quasiparticle branches coined lower (LP) and upper (UP) polaritons. The energy of these states is given by~\cite{Carusotto2013} 
\begin{gather}
\label{pbranch}
\omega_{\text{UP/LP}}(\mathbf k)=\frac{1}{2}\left(\omega_c(\mathbf k)+\omega_X\pm\sqrt{(\omega_c(\mathbf k)-\omega_X)^2+4\Omega^2} \right).   
\end{gather}
Here, $\omega_c(\mathbf k)=\frac{c}{n_c}\sqrt{k_z^2+k_{||}^2}$ considering light  incident along the $z$ axis, perpendicular to the cavity mirrors, with an angle $\theta$, that is $k_z=n_c\omega/c\cos\theta$. The collective coupling of the molecules to light is characterised by the Rabi splitting $\Omega=\sqrt{n}g,$ expressed in terms of the coupling between a single molecule and a cavity photon $g,$ enhanced by the number of molecules. Finally, we introduce the cavity energy detuning from the exciton $\delta(\mathbf k)=\omega_c(\mathbf k)-\omega_X.$ 

By combining energy-momentum microscopy with a slow linear gradient in the active layer thickness we are able to set the energy of the cavity across a wide range that encompasses all relevant detunings (see \cite{SM}). Figure~\ref{Fig1}(c) displays the measured normal incidence reflectance spectrum for varying $\delta=\delta(\mathbf k=0)$.  We observe the appearance of two polariton branches characterised by an avoided crossing with a Rabi coupling of $2\Omega=0.34\text{eV}.$ Here the dashed black curves give the energy of the polariton branches in Eq.~\ref{pbranch}. We estimate  a broadening of the cavity photons of $\gamma_c\approx 0.05~\text{eV},$ and the Lorentzian broadening of the excitons of $\gamma_X\approx 0.065~\text{eV}.$  Being $2\Omega\gg \gamma_X,\gamma_c$, our system is in the strongly interacting regime of light and matter.

Figure.~\ref{Fig1}(c) demonstrates the emergence of two polariton branches exhibiting a striking difference. The LP arises and maintains its well-defined quasiparticle character as the cavity detuning is varied from mostly photonic ($\delta<0$) to excitonic ($\delta>0$). Although the reflectance reduces for large positive detunings, the LP branch remains well-defined and can be interpreted as an ideal polariton, accurately described by Eq.~\ref{pbranch}.

The UP, in contrast, exhibits a more intriguing behaviour. For large and positive $\delta$, a well-defined polariton state emerges with energy in good agreement with Eq.~\ref{pbranch}. Interestingly, as the detuning decreases, the UP becomes blurred until no clear signature of this branch can be identified. Although we expect the signal of the UP to weaken as it evolves from photonic to excitonic, we observe its fading even at maximal mixing ($\delta=0$), which strongly contrasts with the signal of the LP. Contrary to the ideal polaritons that are symmetric at $\delta=0,$ the presence of the matter continuuum makes the UP and LP highly asymmetric. 

The intricate features of the UP can intuitively be understood as follows: The absorption spectrum in Fig.~\ref{Fig1}(b) exhibits a mean exciton peak and a complex enveloping matter continuum for energies larger than that of the bare exciton. Thus, the simple two-level diagram of the exciton acquires a more complex structure, as sketched in Fig.~\ref{Fig1}(a). Coupled to a cavity, the Frenkel excitons hybridise with photons yielding two polariton branches that repel each other in energy. The LP is located at energies lower than the bare exciton one, $\omega_X$, and, thus, it is always pushed away from the matter continuum (see Fig.~\ref{Fig1}(a)). Conversely, the UP lies at energies larger than $\omega_X$. Thus, depending on the cavity detuning, it can be placed within the matter continuum. In Fig.~\ref{Fig1}(c) we signal the matter continuum with dashed white lines, which help in noticing that when the polaritons are driven from large positive to negative detunings, the UP vanishes gradually at its entrance in the continuum. When it lies inside the continuum, it is no longer possible to identify a well-defined polariton, marking the breakdown of the quasiparticle picture. We stress that the white lines in Fig.~\ref{Fig1}(c) are a guide to specify in which range the polariton is ill-defined. However, this is a crossover rather than a sharp transition, that is, the UP evolves smoothly to a well-defined quasiparticle as its energy departs from the continuum.

To support our claims about the validity of the quasiparticle picture discussed above, we appeal to a field theoretical approach which allows studying the polaritons within a general theoretical framework. Let us introduce $\hat x^\dagger$ and $\hat c^\dagger$ the operators that create an exciton and a photon, respectively. The imaginary time Green's function~\cite{Fetter1971}
\begin{gather} 
\mathcal G_{\alpha,\beta}(\mathbf k,\tau)=-\langle T_\tau\{ \hat\psi_{\alpha}(\tau)\hat \psi_{\beta}^\dagger(0)\}\rangle, 
\end{gather}
defined as a matrix, where the indices $\alpha,\beta=\{c,x\}$ give the photon ($\psi_{c}^\dagger(\tau)=\hat c^\dagger(\tau)$) and exciton ($\psi_{x}^\dagger(\tau)=\hat x^\dagger(\tau)$) field operators and $T_\tau$ is the time-ordering operator. Here, we have made the dependence on the incident  wavevector, $\mathbf k$, explicit.

The dynamics of the matrix Green's function is governed by a Dyson's equation $$\mathcal G^{-1}_{\alpha,\beta}(\mathbf k,\omega)=\left[G^{(0)}_{\alpha,\beta}(\mathbf k,\omega)\right]^{-1}-\Sigma_{\alpha,\beta}(\mathbf k,\omega),$$ where  $G^{(0)}_{\alpha,\beta}(\mathbf k,\omega)$ denotes the ideal photonic and excitonic Green's function and it reads as 
$[G^{(0)}_{\alpha,\beta}(\mathbf k,\omega)]^{-1}=
\text{diag}(\omega-\omega_c(\mathbf k), \omega-\omega_X).$
 \begin{figure}[ht]
\centering
    \includegraphics[width=1\columnwidth]{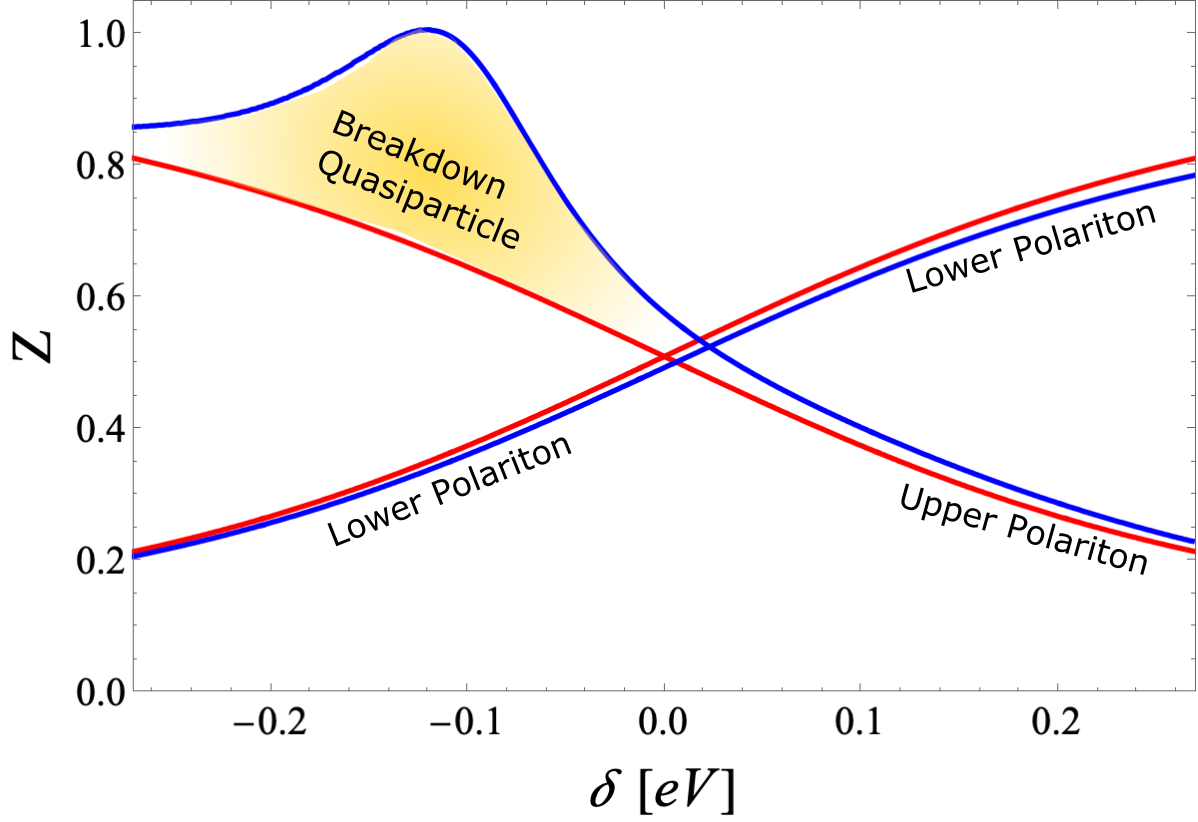}
    \caption{Quasiparticle residue of the polariton branches calculated at normal incidence. Red curves correspond to ideal polaritonsm, while blue ones denote the generalized Hopfield coefficients. The yellow area depicts the deviation from the ideal UP and illustrates the breakdown of the quasiparticle picture for this branch.}
    \label{Fig3}
\end{figure}
 The off-diagonal terms of the self-energy, $\Sigma_{\alpha,\beta}(\mathbf k,\omega)$, are determined by the light-matter coupling $\Sigma_{cx}(\mathbf k,\omega)=\Sigma_{xc}(\mathbf k,\omega)=\Omega,$ whereas the diagonal term $\Sigma_{xx}(\mathbf k,\omega)$ describes the incoherent part of the excitonic spectrum. The polariton energies are calculated by the real poles of the dressed Green's function $\text{Re}\left[\mathcal G^{-1}_{xx}(\mathbf k,\omega)\right]=0$  that account for the light-matter coupling and the matter continuum. In absence of decoherence processes, this equation determines the standard UP/LP polariton energies in Eq.~\ref{pbranch}. While the position of the real poles determines the polaritonic energies, the spectral weight around the poles defines the quasiparticle residue,
\begin{gather}
 Z_{\text{LP(UP)}}(\mathbf k)=\left.\left(\frac{\partial \text{Re}[\mathcal G^{-1}_{xx}(\mathbf k,\omega)]}{\partial \omega}\right)^{-1} \right| _{\omega=\omega_{\text{LP(UP)}}(\mathbf k)}. 
\end{gather}
Within Landau's quasiparticle theory $0<Z_{\text{LP(UP)}}(\mathbf k)\leq 1.$ In absence of incoherent processes, the residue of the quasiparticle is directly connected to the Hopfield coefficients $Z_{LP(UP)}(\mathbf k)=\mathcal C_{\mathbf k}^2 (\mathcal S_{\mathbf k}^2).$  Ideal polaritons retain all the spectral weight in form of coherent excitons, that is,  $Z_{LP}(\mathbf k)+Z_{UP}(\mathbf k)=\mathcal C_{\mathbf k}^2+\mathcal S_{\mathbf k}^2=1$ with  $\mathcal C_{\mathbf k}^2=\frac{1}{2}\left(1+\frac{\delta(\mathbf k)}{\sqrt{\delta^2(\mathbf k)+4\Omega^2}}\right).$

 \begin{figure*}[ht]
\centering
    \includegraphics[width=2\columnwidth]{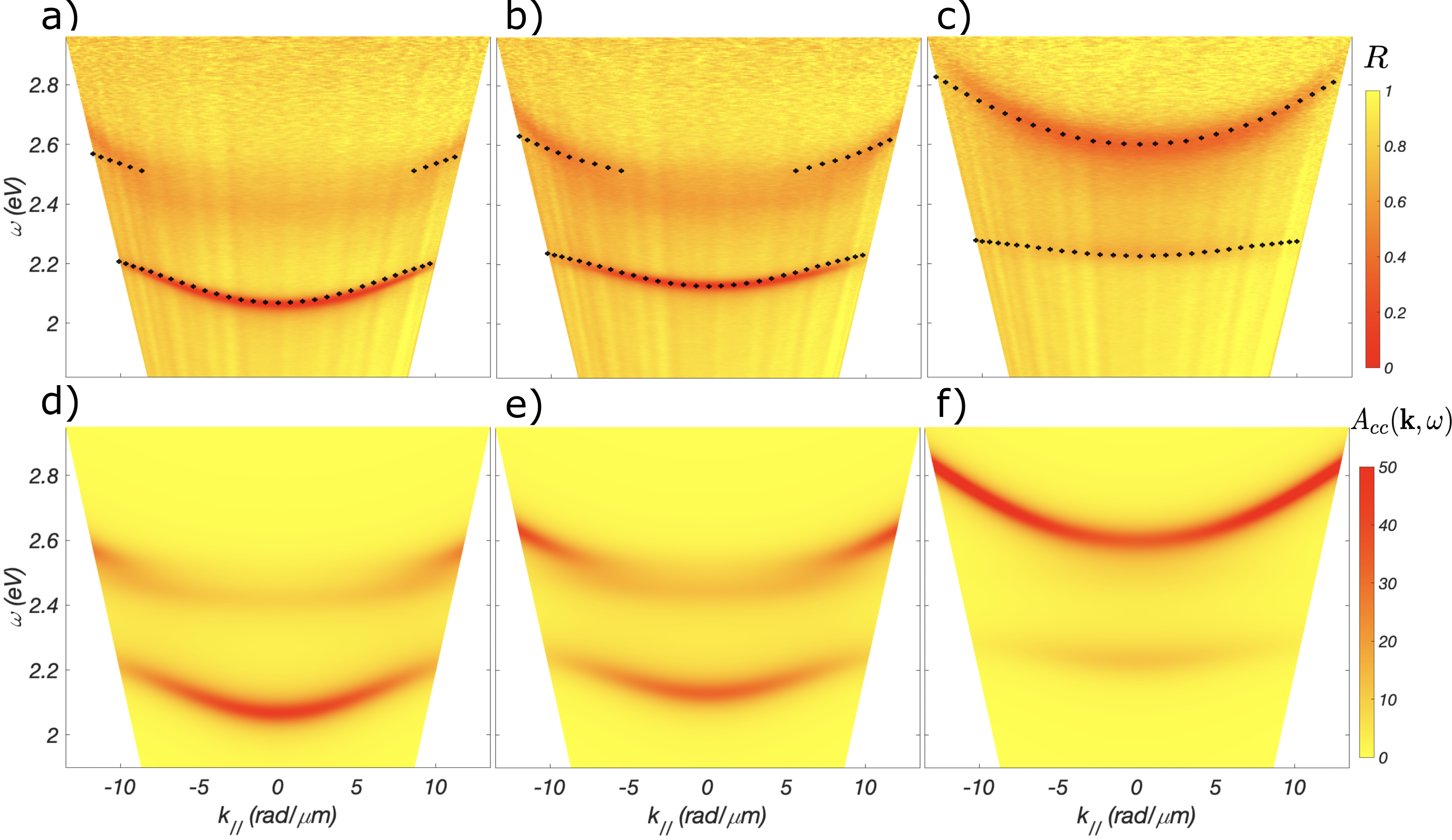}
\caption{(Top) Experimental s-polarized reflectance as a function of the in-plane momentum, $k_{||}$, for several values of the detuning, $\delta$. (a) Negative detuning  $\delta=-0.18\text{eV},$ (b) close to maximal mixing $\delta=-0.07$, and (c) positive detuning $\delta=0.15\text{eV}$. The black squares denote the polariton energies obtained from the field theory. (Bottom) Spectral function of the cavity field for the same parameters than the experiment. }
    \label{Fig2}
\end{figure*}

Our general formalism allows us to explore physics beyond the ideal polaritons. Here, we employ a phenomenological model that closely resembles the experimental absorption in Fig.~\ref{Fig1}(b) (blue curve). Our model contemplates the principal exciton, a weak coupling to a second mode, and Lorentzian broadening of the excitonic and photonic lines as detailed in the SM~\cite{SM}. We start by looking into the residue of the quasiparticle. This defines generalized Hopfield coefficients that may differ from the ideal theory. In Fig.~\ref{Fig3} we show the ideal (red curves) and generalized Hopfield coefficients (blue curves) for the UP and LP. Our theoretical approach obeys the sum-rule $1=\int \frac{d\omega}{2\pi}A_{xx}(\mathbf k,\omega),$ with $A_{xx}(\mathbf k,\omega)=-2\text{Im}\mathcal G_{xx}(\mathbf k,\omega)$, which implies that the spectral weight is always preserved. For large positive and negative detuning, the generalized  Hopfield coefficients asymptotically lead to well-defined quasiparticles for both branches as the matter and light decouple leading to the bare exciton and photon branches. Moreover, the LP is well-described by the ideal theory, for all detunings.  Intriguingly, as the detuning is varied from positive to negative values, we start noticing a stark disagreement between the ideal and generalized Hopfield for the UP. The associated  residue increases and approaches $Z_{UP}(\mathbf k)\sim 1,$ leading to an unphysical value of $Z_{LP}(\mathbf k)+Z_{UP}(\mathbf k)>1$ and, thus, indicating the failing of the quasiparticle approach to describe the UP. We stress that the spectral function remains always normalized for all values of the detuning. The shaded yellow area in Fig.~\ref{Fig3} corresponds to the regime where the sum-rule is conserved but the high-energy branch cannot longer be understood as a  quasiparticle. Therefore, our formalism provides an accurate understanding and gives an interpretation of the involved quantum states beyond the  quasiparticle properties.

It is worth emphasizing that in Fig.~\ref{Fig3} we have added a broadening to the exciton line $\gamma_X=0.065\text{eV}$ corresponding to the absorption spectrum in Fig.~\ref{Fig1}(b) (green curve). This highlights that the standard broadening of the exciton and photon lines does not suffice to explain the breakdown of the quasiparticle approach which arises because of the non-trivial matter spectrum  associated precisely to the {\it bump} missed by the green curve in Fig.~\ref{Fig1}(b).

To further understand the limits of the quasiparticle picture, we now investigate the dispersion of the polariton branches as a function of the in-plane wavevector, $k_{||}$ and for three values of the detuning. For $\delta=-0.18\text{eV}$, shown in Fig.~\ref{Fig2}(a), we observe 
a single well-defined low energy polariton branch and a blurred region at higher energies. Here, the LP arises far from the matter continuum and remains clearly visible for all measured in-plane momentum. The remnant of the UP appears as a foggy region where the reflectance drops. For large values of $k_{||}$, the signatures of the UP are slightly more visible. Here, the black squares correspond to the quasiparticle energies obtained from our field theory. Squares are plotted only in the regions of the dispersion diagram where the ideal quasiparticle picture provides physical results. The success of our extended theory is evident from the excellent agreement found between our measurements and the spectral function of the photon field, $A_{cc}(\mathbf k,\omega)=-2\text{Im}\mathcal G_{cc}(\mathbf k,\omega)$ , plotted in Fig.~\ref{Fig2}(d). 

With increasing $\delta$, as shown for $\delta=-0.07\text{eV}$ in Fig.~\ref{Fig2}(b),  the LP retains its visibility, whereas the UP remains blurred for small wavevectors  and gains coherence for larger ones. This can be understood intuitively: for small wavevectors the polariton is formed predominantly by matter and strongly influenced by the matter continuum. On the contrary, with increasing wavevector the polariton exits the continuum and its linewidth narrows consequence of a dominant photonic component~\cite{SM}. Here again, we find a very good agreement with our theoretical approach as shown in Fig.~\ref{Fig2}(e). 

Finally, in Fig.~\ref{Fig2}(c) we see the emergence of a well-defined UP for positive detuning, $\delta= 0.15\text{eV}$, and for all $k_{||}$, with a very good agreement with the theory shown in Fig.~\ref{Fig2}(f). This is consistent with Fig.~\ref{Fig1}(c) and with our previous discussion. The LP, on the other hand, exhibits an increased reflectance and loses spectral weight. However, the lower visibility of the LP is consequence of the small photonic component rather than a breakdown of the quasiparticle picture for this branch since the matter continuum lies above its energy, as shown in Fig.~\ref{Fig1}.

{\it Conclusions and Outlook.-} Exciton-polaritons provide a simple yet very powerful description of hybrid light-matter systems which have proven to be extremely successful to approach a plethora of systems within the linear and non-linear regime. In this work, we have experimentally unveiled the extent of Landau's quasiparticle theory for organic molecules strongly coupled to cavity photons at room temperature. As result of the molecular absorption, which features a mean exciton peak and an enveloping matter continuum, we have demonstrated the fading of the polariton branch when tuned inside this continuum. We analysed such phenomenon by varying the cavity photon energy and momentum which allowed us to explore intriguing effects in the dispersion diagram. Our conclusions are sustained by a general theoretical framework that expands the study of exciton-polaritons beyond the quasiparticle picture and agrees remarkably well with the experiment. The breakdown of the upper polariton, experimentally observed by the paling and broadening of the reflectance spectrum, is signaled by the inability to associate physical quasiparticle properties to this branch.

Complex intrinsic and/or extrinsic processes are, in general, features that cannot be disentangled from coherent excitations and can render the quasiparticle picture invalid. These sources of decoherence are not necessarily detrimental, but their role can be exploited as a mechanism to detect, probe, and control underlying complicate processes and further be employed to tailor many-body phenomena beyond quasiparticle physics. Finally, their presence may play an important role in the temporal dynamics, scattering thermalization and relaxation of organic exciton-polaritons. Despite being commonly disregarded, the understanding of many-body interactions, as the ones discussed in this Letter, is paramount for advancing the organic polaritons research field.

{\it Acknowledgments.-} G. P. acknowledges financial support from Grants UNAM DGAPA PAPIIT No. IN104522 and CONACyT projects 1564464 and 1098652. H. L. G. acknowledges financial support from Grant UNAM DGAPA PAPIIT No. IA103621. A. C. G. acknowledges financial support from Grant UNAM DGAPA PAPIIT No. IN108620. C. L. O-R acknowledges financial support from Grant UNAM DGAPA PAPIIT IG100521.

\begin{widetext}
    
\appendix

\section{Sample fabrication and characterization}

The Fabry–P\'erot cavities have been fabricated on glass substrates $10\times 10 \text{mm}^2$  cleaned by sonication with ethanol, water and ethanol for 10min each. The cleaned glass substrates were dried with N$_2$ and then left in the oven overnight to evaporate residues of solvent and water. Immediately before the thin film deposition, the substrates were plasma-cleaned, with air as the plasma source, for 5 min with a Harrick Plasma cleaner. A solution of polyvinyl alcohol (PVA, Mowiol 44-88, 86.7-88.7$\%$ hydrolyzed, Mw $\approx$ 205 000 g/mol) in distilled water was prepared by adding 25 mg of PVA to 1 mL of distilled water. The solution was stirred at 70$^{\circ}$C for 24 hours to ensure the complete dissolution of the PVA. Then, 9.5 mg of Erythrosin B (EryB, Sigma Aldrich with dye content $>$90$\%$) was added to the PVA/water solution and left stirring at room temperature for 5 hours, which yielded a 0.5 M concentration of EryB in PVA. The EryB/PVA thin films were deposited by spin-coating at 2600 rpm using a 0.45 $\mu$m pore PTFE syringe filter, obtaining approximately 160 nm thickness. Finally, the thin film was annealed at 75$^{\circ}$C for 15 min.

Ag mirrors were fabricated by magnetron sputtering technique at room temperature. The sputtering chamber was evacuated to a base pressure of approximately 10$^{-6}$ Torr and then was pressurized with argon flow to 3 x 10$^{-3}$ Torr during deposition. We used 2-inch diameter silver (Ag) target at 99.99$\%$ purity with power of 25 W and a deposition rate of 0.19 nm/s. A thick 300 nm Ag film was sputtered on top of the glass substrate and the semitransparent 25 nm Ag film was sputtered on top of the active layer.



\section{Experimental set-up}
Energy-momentum spectroscopy is performed in a homemade confocal Fourier optical microscope. It consists in imaging the back-focal plane of a high numerical aperture microscope objective onto the entrance slit of a spectrograph (Andor Kymera 328i) coupled to a sCMOS camera (Andor Zyla 4.2P). This is done by using a Bertrand lens which provides direct access to the angular- and spectral-resolved reflectance. In our set-up the sample is illuminated through a Plan Fluor 100x/0.9 NA objective with white light emitted by a halogen lamp. The focal spot full-width at half-maximum equals 7 $\mu$m. The collected light is dispersed by a diffraction grating (150 lines mm, blazed at 500 nm). Two linear polarizers in the excitation and collection path are used to select the s- or p- polarization. Angular-resolved reflectance is obtained by replacing the cavity with a commercial mirror, which allows to normalize the spectra reflected off the cavity at each angle with those obtained with the mirror at the corresponding angles.

\section{Detuning-resolved measurements}
Experimentally, it represents a challenge to tune on demand the energy of the cavity. To overcome this problem, we device the active layer such as it is  homogeneous at the center but feature a very slow and almost linear gradient towards the peripheral zones of the sample. This gradient is observed radially from the center of the cavity and it is a direct consequence of the spin-coating used to fabricate the active layer. The full range of detuning is obtained by shifting the incident beam by approximately one millimiter and corresponds to approximately 30 nm increase on the nominal thickness. The position of the focal spot is controlled by micrometer screws that permit shifting the focal spot along the gradient. This provides us access to all relevant detunings.

\section{Exciton-Polaritons: Theory}
In absence of incoherent processes the Hamiltonian describing exciton-polaritons is given by
\begin{gather}
\hat H=\omega_c\hat a^\dagger\hat a+\omega_X \hat x^\dagger\hat x+\Omega(\hat a^\dagger\hat x+\hat x^\dagger\hat a),    
\end{gather}
where $\hat a^\dagger$ and $\hat x^\dagger$ are the operators that create a cavity photon and an exciton with energy $\omega_c(\mathbf k)$ and $\omega_X$, respectively. Here, $\omega_c(\mathbf k)$ depends on the incident parallel wavevector, which in turns depends on the angle of incidence. The light-matter coupling $\Omega$ incorporates the collective coupling between the two-level emitters and the cavity field $\Omega=\sqrt{n}g$ where $g$ is the single photon coupling and $n$ the concentration. The Hamiltonian can be written in terms of its eigenvectors coined upper and lower polaritons, linear combinations of the cavity photons and excitons 
\begin{gather}
 \hat L(\mathbf k)=\mathcal C_{\mathbf k}\hat x+\mathcal S_{\mathbf k}\hat a \\ \nonumber
 \hat U(\mathbf k)=-\mathcal S_{\mathbf k}\hat x+\hat C_{\mathbf k}\hat a,
\end{gather}
which introduces the standard Hopfield coefficients
\begin{gather}
 \mathcal C_{\mathbf k}=\sqrt{\frac{1}{2}\left(1+\frac{\delta(\mathbf k)}{\sqrt{\delta^2(\mathbf k)+4\Omega^2}}\right)}   \\ \nonumber
\mathcal S_{\mathbf k}= \sqrt{\frac{1}{2}\left(1-\frac{\delta(\mathbf k)}{\sqrt{\delta^2(\mathbf k)4\Omega^2}}\right)}.
\end{gather}
This approach is extremely useful to characterise the polaritons within the single particle picture. To systematically include additional effects we employ a field theoretical approach based on the imaginary time Green's function
\begin{gather} 
\mathcal G_{\alpha,\beta}(\tau)=-\langle T_\tau\{ \hat\psi_{\alpha}(\tau)\hat \psi_{\beta}^\dagger(0)\}\rangle, 
\end{gather}
 defined as a matrix, where the indices $\alpha=\{c,x\}$ correspond to the photon and exciton fields respectively. 

 The Green's function is governed by a Dyson's equation 
 
\begin{gather}
\mathcal G^{-1}_{\alpha,\beta}(\mathbf k,\omega)=G^{(0)}_{\alpha,\beta}(\mathbf k,\omega)^{-1}-\Sigma_{\alpha,\beta}(\mathbf k,\omega),
\end{gather} 
here $G^{(0)}_{\alpha,\beta}(\mathbf k,\omega)$ denotes the ideal photonic and excitonic Green's function
\begin{gather}
G^{(0)}(\mathbf k,\omega)=
 \begin{bmatrix}
     \frac{1}{\omega-\omega_c(\mathbf k)} & 0 \\
     0 &\frac{1}{\omega-\omega_X}
 \end{bmatrix},
\end{gather}
on the other hand, the off-diagonal terms of the self-energy are determined by the light-matter coupling
\begin{gather}
\Sigma(\mathbf k,\omega)=
 \begin{bmatrix}
     0 & \Omega \\
     \Omega &\Sigma_{xx}(\mathbf k,\omega)
 \end{bmatrix},
\end{gather}
whereas the diagonal term $\Sigma_{xx}(\mathbf k,\omega)$ accounts for the incoherent part of the excitonic spectrum. 
\subsection{Ideal Polaritons}
For $\Sigma_{xx}(\omega)=0$ the ideal polaritons emerge as a poles of the dressed Green's function, that is
\begin{gather}
\text{Re}\left[\mathcal G^{-1}_{xx}(\mathbf k,\omega)\right]=0,  
\end{gather}
which has two solutions simply given by the polaritonic energies
\begin{gather}
\label{pbranch}
\omega_{\text{UP/LP}}(\mathbf k)=\frac{1}{2}\left(\omega_c(\mathbf k)+\omega_X\pm\sqrt{(\omega_c(\mathbf k)-\omega_X)^2+4\Omega^2} \right).   
\end{gather}
The residue of the poles 
 \begin{gather}
 Z_{\text{LP/UP}}(\mathbf k)=\left.\frac{1}{\frac{\partial \text{Re}[\mathcal G^{-1}_{xx}(\mathbf k,\omega)]}{\partial \omega}} \right| _{\omega=\omega_{\text{LP/UP}}(\mathbf k)}, 
\end{gather}
relates to the Hopfield coefficients $\mathcal C_{\mathbf k}^2=Z_{\text{LP}}(\mathbf k)$ and $\mathcal S_{\mathbf k}^2=Z_{\text{UP}}(\mathbf k)$. We introduce the spectral function for the excitons and photons
\begin{gather}
A_{\alpha\alpha}(\mathbf k,\omega)=-2\text{Im}\mathcal G_{\alpha\alpha}(\mathbf k,\omega),
\end{gather}
for a system with solely coherent excitations (quasiparticles), we have that the whole spectral weight is distributed in the real poles, for the excitons it reads 
\begin{gather}
 \mathcal G_{xx}(\omega)=\frac{Z_{\text{LP}}}{\omega-\omega_{\text{LP}}}+\frac{Z_{\text{UP}}}{\omega-\omega_{\text{UP}}}, 
\end{gather}
thus,
\begin{gather}
A_{xx}(\mathbf k,\omega)=2\pi\left(Z_{\text{LP}}(\mathbf k)\delta(\omega-\omega_{\text{LP}}(\mathbf k))+Z_{\text{UP}}(\mathbf k)\delta(\omega-\omega_{\text{UP}}(\mathbf k))\right),
\end{gather}
and we have 
\begin{gather}
\int \frac{d\omega}{2\pi}A_{xx}(\mathbf k,\omega)=1=Z_{\text{LP}}(\mathbf k)+Z_{\text{UP}}(\mathbf k),   
\end{gather}
that is, ideal polaritons are formed only by coherent excitations.
\subsection{Polaritons in the continuum}
Our theory conserves always the spectral weight, that is, 
\begin{gather}
\label{sum}
    \int \frac{d\omega}{2\pi}A_{xx}(\mathbf k,\omega)=1,
\end{gather} remains normalized. However, in the presence of incoherent processes it may not be longer true that the spectral function is formed only be coherent quasiparticle. In general, the incoherence arises from complicated intrinsic processes which microscopic description lies beyond the scope of the manuscript. Here, we adopt an experimental-driven approach where we include these processes at a more phenomenological level. First, we consider a Lorentzian broadening of the cavity and main exciton peak, this can be done by adding an imaginary width to the cavity and exciton field. In addition, we include coupling between the cavity field and additional vibrational modes.
\begin{gather}
 \left[\mathcal G_{cc}(\mathbf k,\omega)\right]^{-1}=\omega-\omega_c(\mathbf k)+i\gamma_c-\frac{\Omega^2 Z_{X}}{\omega-\omega_X+i\gamma_X}-\frac{\Omega^2 Z_{h}}{\omega-\omega_{h}+i\gamma_{h}},
\end{gather}
that is, we model the excitonic Green's function in terms of the bare excitonic transition coupled to a single vibrational mode
\begin{gather}
[G^{0}_{xx}(\omega)]^{-1}=\frac{ Z_{X}}{\omega-\omega_X+i\gamma_X}-\frac{ Z_{h}}{\omega-\omega_{h}+i\gamma_{h}}    
\end{gather}
with $Z_{X}+ Z_{h}=1$ such that the light-matter coupling is distributed in these two transitions. 

This approach preserves the sum rule in Eq.~\ref{sum} which we also verified numerically, allows us to mimic the {\it bump} and the broadening of the exciton and cavity fields while restricting to the minimum the number of fitting parameters. 

In Fig. 1(b) of the main text we compare the absorption obtained from our formalism against the experimental  measurement. We observe that our approach fits reasonably well the experiment. For the theory we take a Rabi splitting of $2\Omega=0.34\text{eV},$ which splits into $Z_{X}=0.9,$ excitonic energies of $\omega_X=2.33\text{eV},$ $\omega_{h}=\omega_X+\delta\omega$ with $\delta\omega/\omega_X=0.094,$ $\gamma_X=0.067\text{eV}$ and $\gamma_{h}=0.065\text{eV}.$  We stress that the broadening of the main exciton mode overlaps with the position of the bump, thus, the full description beyond the single two-level picture for the exciton is required. Finally, the breakdown of the quasiparticle picture can be accompanied by a difference of the quasiparticle energies and residues calculated using the dressed excitonic or photonic Green's function.

To supplement our discussion in the main text, in Fig.~\ref{Fig4N} we calculate the quasiparticle residues for the quasiparticles branches shown in the main text. For clarity and presentation purposes we show again in Fig.~\ref{Fig4N}(a)-(c) the experimental s-polarized reflectance as a function of the incident wavevector for the same values of detuning as in Fig. 3 of the main text. In Fig.~\ref{Fig4N} (d)-(f) we show the corresponding quasiparticle residue. Here, the blue dots correspond to the quasiparticle residue for ideal polaritons, while the red dots give the generalized Hopfield coefficients. As discussed in the main text, we find a remarkable agreement between ideal and generalized polaritons for the lower branch. The upper polariton exhibits more structure and can largely deviate from the ideal polaritons, in particular, in the main text we only plotted the energies of the upper polariton where the quasiparticle picture remains an accurate description of the system. As a further comment on the validity of our theory, we highlight that the breakdown of the quasiparticle picture is signalled by a sudden linewidth broadening along the upper polariton dispersion, which roughly corresponds to the crossing point of the generalized Hopfield coefficients. Conversely, the crossing point of the ideal Hopfield coefficients fails to predict the wavevector at which the broadening appears. The broadening found at small wavevectors is explained by a dominant matter component in the polaritons, which is strongly influenced by the matter continuum. On the contrary, for large wavevectors the predominantly photonic character of the polaritons establishes a higher degree of coherence and, thus, a narrower dispersion.

\begin{figure}[h!]
\centering
    \includegraphics[width=.96\columnwidth]{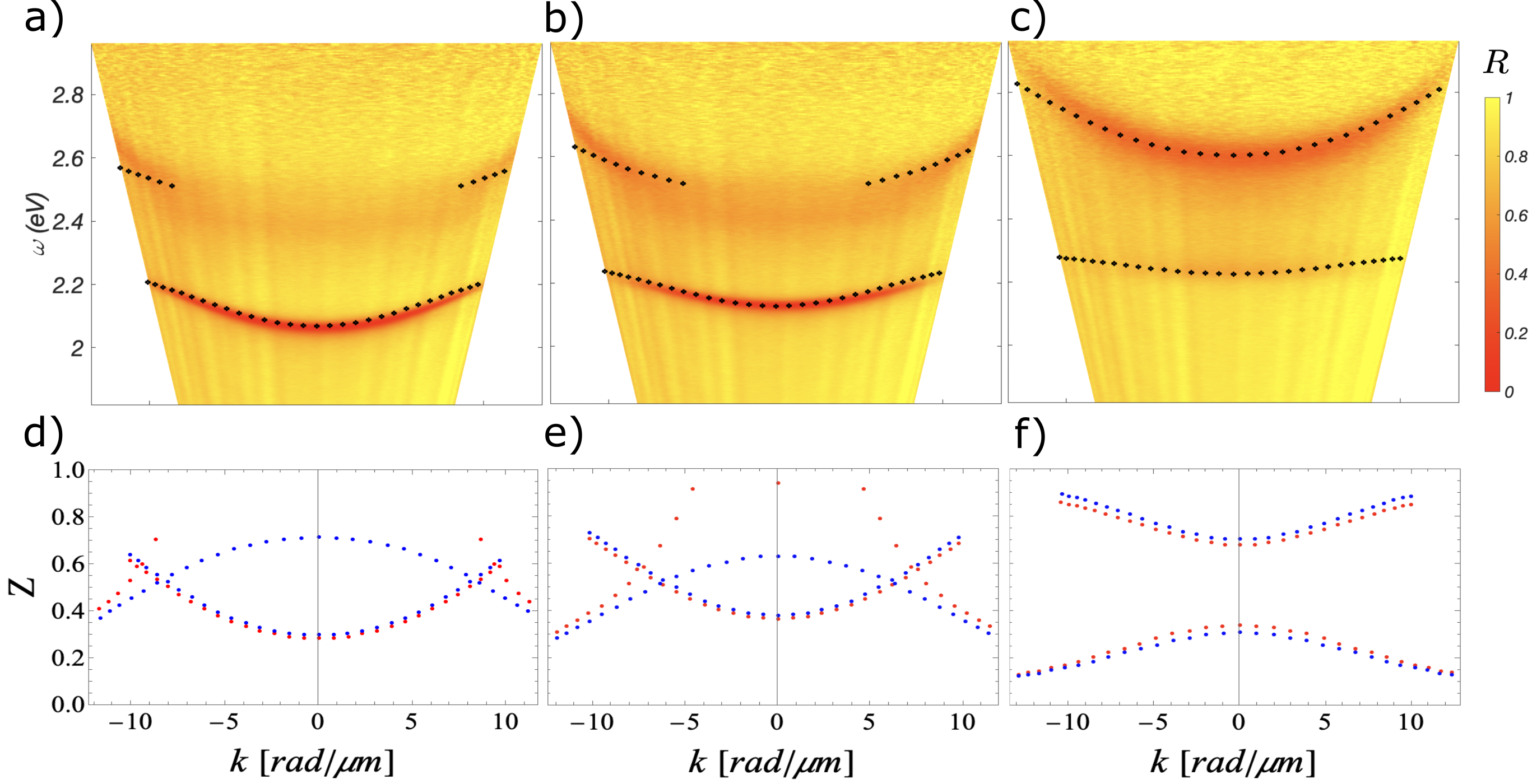}
    \caption{(a)-(c) Experimental s-polarized reflectance as a function of the in-plane momentum for the same values of detuning as in Fig. 3 of the main text. (d)-(f) Quasiparticle residue corresponding to (a)-(c). The blue dots correspond to the ideal polaritons, the red dots to the generalized Hopfield coefficients. 
    } 
    \label{Fig4N}
\end{figure}

\end{widetext}
\bibliography{references} 
\end{document}